\documentclass[12pt]{article}
\catcode`\@=11
\@addtoreset{equation}{section}

\global\arraycolsep=2pt 
\newcommand{\ga}{\gamma}

\newcommand{\la}{\lambda}

\newcommand{\Tr}{{\rm Tr}}

\newcommand{\alg}[1]{\mathfrak{#1}}

\newcommand{\nln}{\nonumber\\}

\newcommand{\mL}{\mathcal L}

\newcommand*{\AdS}[1]{\ensuremath{\text{AdS}_{#1}}}

\usepackage{mathrsfs,amsbsy,amssymb,latexsym,amsfonts,amsmath,cite}
\usepackage{graphicx,color}
\usepackage{mathtools}

\allowdisplaybreaks

\begin{document}

\begin{flushright}
\parbox{4cm}
{KUNS-2578
\\ \today
} 
\end{flushright}

\vspace*{1cm}

\begin{center}
{\Large \bf 
Yang-Baxter sigma models and Lax pairs \\ arising from $\kappa$-Poincar\'e $r$-matrices
}
\vspace*{0.75cm}\\
{\large Andrzej Borowiec$^{\ast}$\footnote{E-mail:~andrzej.borowiec@ift.uni.wroc.pl},
Hideki Kyono$^{\dagger}$\footnote{E-mail:~h\_kyono@gauge.scphys.kyoto-u.ac.jp}, 
Jerzy Lukierski$^{\ast}$\footnote{E-mail:~jerzy.lukierski@ift.uni.wroc.pl}, \\ 
Jun-ichi Sakamoto$^{\dagger}$\footnote{E-mail:~sakajun@gauge.scphys.kyoto-u.ac.jp}, 
and Kentaroh Yoshida$^{\dagger}$\footnote{E-mail:~kyoshida@gauge.scphys.kyoto-u.ac.jp}} 
\end{center}

\vspace*{0.25cm}

\begin{center}
$^{\ast}${\it Institute for Theoretical Physics, University of Wroclaw, \\
pl.\ Maxa Borna 9, 50-206 Wroclaw, Poland} 
\vspace*{0.25cm}\\
$^{\dagger}${\it Department of Physics, Kyoto University, \\ 
Kitashirakawa Oiwake-cho, Kyoto 606-8502, Japan} 
\end{center}

\vspace{1cm}

\begin{abstract}
We study Yang-Baxter sigma models with deformed 4D Minkowski spacetimes 
arising from classical $r$-matrices 
associated with $\kappa$-deformations of the Poincar\'e algebra. 
These classical $\kappa$-Poincar\'e $r$-matrices describe three kinds of deformations:  
1) the standard deformation, 2) the tachyonic deformation, and 3) the light-cone deformation. 
For each deformation, the metric and two-form $B$-field are computed 
from the associated $r$-matrix. 
The first two deformations, related to the modified classical Yang-Baxter equation,  
lead to T-duals of dS$_4$ and AdS$_4$\,, respectively. 
The third deformation, associated with the homogeneous classical Yang-Baxter equation,  
leads to a time-dependent pp-wave background. 
Finally, we construct a Lax pair for the generalized $\kappa$-Poincar\'e $r$-matrix 
that unifies the three kinds of deformations mentioned above as special cases. 
\end{abstract}

\setcounter{footnote}{0}
\setcounter{page}{0}
\thispagestyle{empty}

\newpage

\tableofcontents

\section{Introduction}

A fascinating topic in the field of integrable systems is 
a systematic way to study integrable deformations of 2D non-linear sigma models. 
It was proposed by Klimcik \cite{Klimcik} originally for principal chiral models 
as a higher-dimensional generalization of 3D squashed sphere\footnote{
The classical integrability of the squashed S$^3$ was discussed 
in classic papers \cite{Cherednik,FR,BFP}.  
For related Yangians and quantum affine algebras, see \cite{KY,KOY,KYhybrid,KMY-QAA,ORU,DMV-WZW}. 
For an earlier attempt towards the higher-dimensional case, see \cite{BR}. 
}. Following this method, integrable deformations are provided by classical $r$-matrices  
which satisfy the modified classical Yang-Baxter equation (mCYBE)\,. 
Then, it is so powerful that, given a classical $r$-matrix, 
the associated Lax pair for principal chiral models follows automatically! 

\medskip 
  
Klimcik's method was subsequently generalized to symmetric cosets \cite{DMV}. 
The AdS$_5\times$S$^5$ manifold is an example of a symmetric coset, 
and in fact the Green-Schwarz action of type IIB superstring on this background
can be constructed on the following supercoset \cite{MT}
\begin{eqnarray}
\frac{PSU(2,2|4)}{SO(1,4)\times SO(5)}\,.  
\end{eqnarray}
Then the $\mathbb{Z}_4$-grading property of this supercoset ensures 
classical integrability \cite{BPR}. Thus, Delduc-Magro-Vicedo presented 
a Yang-Baxter $q$-deformation of the AdS$_5\times$S$^5$ superstring \cite{DMV2}, 
generated by the $r$-matrix of Drinfeld-Jimbo type \cite{DJ}. 

\medskip 

Furthermore, one may consider deformations linked with the homogeneous 
classical Yang-Baxter equation (CYBE)\,. 
The deformed action of the AdS$_5\times$S$^5$ superstring 
based on the CYBE was constructed in \cite{KMY-Jordanian-typeIIB} 
(for purely bosonic sigma models, see \cite{MY-YBE,KY-Sch,Kame})\,. 
In a series of works \cite{LM-MY,MR-MY,Sch-MY,SUGRA-KMY,MY-duality,Stijn1,Stijn2} 
various classical $r$-matrices have been identified with 
type IIB supergravity solutions like Lunin-Maldacena-Frolov backgrounds \cite{LM,Frolov}, 
gravity duals for noncommutative gauge theories \cite{HI,MR}, 
and Schr\"odinger spacetimes \cite{MMT}. 
In particular, string theories on these backgrounds are equivalent to 
the undeformed theories with twisted boundary conditions \cite{Frolov,AAF,KKSY}\footnote{
For general discussions on Jordanian twists and twisted periodic boundary conditions, 
see \cite{Benoit,Stijn2,KY-Sch}.}.  
This relation between a class of gravity solutions and classical $r$-matrices 
has been called the gravity/CYBE correspondence \cite{LM-MY} 
(For a short summary, see \cite{MY-summary}). 
To establish this conjectured relation, there remain many issues to be studied. 

\medskip 

It is worth noting that the gravity/CYBE correspondence may work beyond the integrability. 
A landmark is the AdS$_5\times T^{1,1}$ background, where $T^{1,1}$ is a Sasaki-Einstein manifold.  
On this background, classical string solutions exhibit chaotic motions \cite{BZ} 
and hence the complete integrability is broken. 
On the other hand, TsT transformations can be performed for $T^{1,1}$ \cite{LM,CO} and 
give rise to (non-integrable) deformations of $T^{1,1}$\,. 
Interestingly enough, these deformations can be reproduced as Yang-Baxter 
deformations with abelian classical $r$-matrices \cite{CMY}. This result indicates that 
the Yang-Baxter deformations technique may work as well for non-integrable cases.  

\medskip 

Based on this success, it would be interesting to study Yang-Baxter deformations 
of 4D Minkowski spacetime \cite{YB-Min}. 
This may be regarded as an application of the gravity/CYBE correspondence to the case of flat space. 
It is the fundamental question to reveal to what extent the correspondence can be generalized.  
Furthermore, in comparison to curved spaces like AdS spaces, one can readily see the relation between 
Yang-Baxter deformations and duality transformations like T-duality and S-duality. Further it would be 
helpful to study non-perturbative aspects of string theory and new ideas such as non-geometric backgrounds 
and doubled geometries. 
Hence it may be possible to figure out a connection between Yang-Baxter deformations and quantum spectra. 
As a matter of course, it is quite significant to unveil quantum aspects of Yang-Baxter deformations. 

\medskip

However, in the case of Minkowski spacetime, 
there is an obstacle that the inner product entering into the YB sigma model action 
is degenerate. A possible way around is to employ an embedding of 4D Minkowski spacetime into the bulk AdS$_5$\,. 
By adopting this resolution\footnote{To employ this resolution, we are now considering the 4D case, 
though it is also interesting to consider the deformations of 10D Minkowski spacetime 
as string target space.},
 a Yang-Baxter sigma model providing deformations of 4D Minkowski spacetime was proposed 
in \cite{YB-Min}. Then, classical $r$-matrices have been identified 
with a large number of gravity solutions 
such as Melvin backgrounds\cite{Melvin,GM,Tseytlin,HT}, pp-wave backgrounds \cite{NW}, 
Hashimoto-Sethi backgrounds \cite{HS} and Spradlin-Takayanagi-Volovich backgrounds \cite{STV}. 
More interestingly, T-duals of dS$_4$ and AdS$_4$ also have been reproduced 
from classical $r$-matrices\footnote{The appearance of dS space is originally discussed in \cite{Klimcik}. 
It is worth noting that (T-duals of) dS spaces appear in various contexts \cite{DMV2,HRT,AT}.}. 
In addition, new backgrounds generated by the standard $q$-deformation were also presented. 
In comparison to the Yang-Baxter deformations of principal chiral models and 
the usual symmetric cosets, the general form of Lax pair has not been constructed yet. 
But the solvability or integrability of all the models listed above, apart from the new ones, 
has already been shown. It would be an interesting task to show the integrability 
for a sufficiently large class of Yang-Baxter deformations.

\medskip 

On the other hand, there is a long history of studies describing the quantum deformations 
of Poincar\'e algebras, starting from \cite{kappa}, and subsequent studies 
of quantum-deformed conformal algebras 
\cite{JN,KJ,LW}. Deformed Poincar\'e algebras are classified in terms of 
classical $r$-matrices similarly as the Yang-Baxter deformations (for still incomplete list of 
$D=4$ Poincar\'e $r$-matrices, depending only on the Poincar\'e generators, see \cite{Z,Tolstoy}). 
It is of particular interest to study a special class of $r$-matrices 
generating the $\kappa$-deformations of the Poincar\'e algebra. 
 
\medskip 

In this paper, we will study Yang-Baxter sigma models 
with deformed 4D Minkowski spacetimes arising from three classical $r$-matrices 
associated with three $\kappa$-deformations of the Poincar\'e algebra. 
The classical $r$-matrices include three kinds of deformations: 
1) the standard deformation (non-split type)\,, 
2) the tachyonic deformation (split type)\,, 
and 3) the light-cone deformation (homogeneous type)\,. 
For each type of deformation, the metric and two-form $B$-field are computed. 
The first two cases are described by the mCYBE 
and lead to T-duals of dS$_4$ and AdS$_4$\,, respectively. 
The third case is linked with the CYBE 
and gives rise to a time-dependent pp-wave background. 
Then we construct the deformed Lax pairs for each of the cases. 
All of the resulting backgrounds are classically integrable 
and one can derive the corresponding Lax pairs. 
In fact, we have succeeded in constructing a Lax pair 
for the generalized $\kappa$-Poincar\'e $r$-matrix 
depending on the constant four vector $a_\mu$ 
that unifies the three deformations as special cases. 
The unified Lax pair interpolates between the three cases 
and remarkably the three different classes of $\kappa$-deformations 
are treated by one set of equations.

\medskip 

Before discussing our results, we shall comment on the choice of dimension D=4 used in this paper. 
D=4 string theory is noncritical and it is described by a 2D Liouville gravity \cite{Polyakov}. 
Our $\kappa$-deformed D=4 Yang-Baxter sigma models can be considered as a preliminary stage 
for the description of $\kappa$-deformed D=4 noncritical string 
with the 2D Liouville gravity action 
representing an extension of the sigma-model approach to a string description. 
Such a ``noncritical'' extension of Yang-Baxter sigma models 
still should be formulated, however because the D=2 Liouville gravity is integrable 
(For a Lax pair in W-gravity, see \cite{Aoki}), it is plausible, 
in the context of proven integrability in this paper, that the integrability should remain valid 
after performing the $\kappa$-deformation. 

\medskip 

This paper is organized as follows. 
In section 2 we revisit coset constructions of AdS$_5$\,. 
In section 3 we provide a short review of the Yang-Baxter deformation method 
and its adjustment to 4D Minkowski spacetime. 
In section 4 we recall the classical $r$-matrices 
defining three types of $\kappa$-Poincar\'e algebras. 
Furthermore, the deformed metric and two-form $B$-field are computed 
for the three $\kappa$-deformations.   
In section 5 the Lax pair is constructed 
for the general $a_\mu$-dependent $\kappa$-Poincar\'e $r$-matrix 
and then the concrete formulae for Lax pairs for the three cases are given. 
In section 6 we present discussion and final remarks. 
In Appendix A, a connection among coset representatives of AdS$_5$ is presented. 
In Appendix B, we argue another coset for the Poincar\'e AdS$_5$\,. 
In Appendix C, we clarify the origin of the coincidence of target spaces 
derived via Yang-Baxter sigma model from different classical $r$-matrices. 

\paragraph{NOTE:} When preparing our draft, an interesting paper \cite{PT} has appeared 
on the arXiv. The present paper and \cite{PT} have some overlapping 
on a T-dual of dS$_4$ for the standard $\kappa$-deformation.

\section{Revisiting coset constructions of AdS$_5$}

\subsection{4D conformal algebra $\alg{so}(2,4)$ and its spinorial realization}

In the first place, we shall introduce a fundamental spinorial representation 
of the Lie algebra $\mathfrak{so}(2,4) \cong \mathfrak{su}(2,2)$\,. 
The generators $J_{ab}~(a,b=0,1,2,3,4,5)$ satisfy the following relations:
\begin{eqnarray}
[J_{ab},J_{cd}] = \eta_{ad}J_{bc}+\eta_{bc}J_{ad}-\eta_{ac}J_{bd}-\eta_{bd}J_{ac}\,.
\end{eqnarray}
Here $\eta_{ab}=(-1,1,1,1,-1,1)$\,. 

\medskip 

Let us introduce the gamma matrices $\gamma_{\mu}~(\mu=0,1,2,3)$ and $\gamma_5$ 
satisfying the relations: 
\begin{eqnarray}
\{\gamma_{\mu},\gamma_{\nu}\} = 2\eta_{\mu\nu}\,, \quad 
\gamma_5 \equiv -i\gamma_0\gamma_1\gamma_2\gamma_3\,. \label{dirac}
\end{eqnarray}
Then, by introducing the following quantities, 
\begin{equation}
n_{\mu\nu} \equiv \tfrac{1}{4}[\ga_\mu,\ga_\nu]\,, \qquad 
n_{\mu5} \equiv \tfrac{1}{4}[\ga_\mu,\ga_5]\,, 
\end{equation}
one can introduce a spinor representation\footnote{The generators in (\ref{2.4}) generate the spinorial 
double covering $SU(2,2)$ of $SO(2,4)$\,, which only matters in the case of global (topological) considerations.  
The spinor representation is rather convenient.} 
of $\alg{so}(2,4)$ 
\begin{eqnarray}
J_{\mu\nu} = n_{\mu\nu}\,, \quad J_{\mu 4} = \frac{1}{2}\gamma_{\mu}\,,  
\quad J_{\mu 5} = n_{\mu 5}\,, \quad J_{54} = \frac{1}{2}\gamma_5\,.  \label{2.4}
\end{eqnarray}
Thus the Lie algebra $\alg{so}(2,4)$ 
and a subalgebra $\alg{so}(1,4)$ are represented by 
\begin{align}
\alg{so}(2,4)&= {\rm span}_{\mathbb{R}}\{~\frac{1}{2}\ga_\mu\,,\frac{1}{2}\ga_5\,,
n_{\mu\nu}\,, n_{\mu5}~|~~\mu,\nu=0,1,2,3~
\}\,, \label{24} \\ 
\alg{so}(1,4)&= {\rm span}_{\mathbb{R}}\{~
n_{\mu\nu}\,, n_{\mu5}~|~~\mu,\nu=0,1,2,3~\}\,. \label{13}
\end{align}
We obtain the matrix realization of 4D conformal algebra if we substitute the following realization of 
the algebra (\ref{dirac})
\cite{AF,KMY-Jordanian-typeIIB}: 
\begin{align}
&\ga_1=\begin{pmatrix}
0&~0~&~0~&-1\\
0&0&1&0\\
0&1&0&0\\
-1&0&0&0
\end{pmatrix}\,,
\quad
\ga_2=
\begin{pmatrix}
0&~0~&~0~&i\;\\
0&0&i&0\\
0&-i&0&0\\
-i&0&0&0 
\end{pmatrix}\,, 
\quad 
\ga_3=
\begin{pmatrix}
0~&~0~&~1~&~0\\
0&0&0&1\\
1&0&0&0\\
0&1&0&0
\end{pmatrix}\,, 
\nln  
& \ga_0=
\begin{pmatrix}
0&~0~&~1~&0\\
0&0&0&-1\\
-1&0&0&0\\
0&1&0&0
\end{pmatrix}\,, 
\quad 
\ga_5= 
\begin{pmatrix}
\;1~&~0~&0&0\\
0&1&0&0\\
0&0&-1&0\\
0&0&0&-1
\end{pmatrix}\,.
\end{align}

\medskip 

Then we introduce a conformal basis for $\mathfrak{so}(2,4)$ as follows: 
\begin{eqnarray}
\mathfrak{so}(2,4) = {\rm span}_{\mathbb{R}}\{~p_{\mu}\,, n_{\mu\nu}\,, \hat{d}\,, k_{\mu}~|
~~\mu,\nu=0,1,2,3~\}\,. \label{cb}
\end{eqnarray}
Here the translation generator $p_{\mu}$\,, the dilatation $\hat{d}$  
and the special conformal generator $k_{\mu}$ 
are represented by, respectively,  
\begin{align}
p_{\mu} \equiv \frac{1}{2}(\ga_{\mu}-2n_{\mu5})\,, \quad 
\hat{d} \equiv \frac{1}{2}\gamma_5\,,  
\quad 
k_{\mu} \equiv \frac{1}{2}(\ga_{\mu}+2n_{\mu5})\,. 
\end{align}
The non-vanishing commutation relations are given by 
\begin{eqnarray}
[p_\mu ,k_\nu]&=& 2 (n_{\mu\nu}+\eta_{\mu\nu}\, \hat{d}\,)\,,\quad 
[\hat{d},p_{\mu}]=p_\mu\,,\quad [\hat{d},k_\mu]=-k_\mu\,,\nonumber\\
\left[p_\mu,n_{\nu\rho}\right] &=&\eta_{\mu\nu}\, p_\rho-\eta_{\mu\rho}\, p_\nu \,,\quad 
[k_\mu,n_{\nu\rho}]=\eta_{\mu\nu}\,k_\rho-\eta_{\mu\rho}\,k_\nu\,, \nonumber\\
\left[n_{\mu\nu} ,n_{\rho\sigma}\right]&=&\eta_{\mu\sigma}\, n_{\nu\rho}
+\eta_{\nu\rho}\,n_{\mu\sigma}-\eta_{\mu\rho}\,n_{\nu\sigma}-\eta_{\nu\sigma}\,n_{\mu\rho}\,.
\end{eqnarray}

\subsection{Coset construction of the global AdS$_5$}

The AdS$_5$ space is homogeneous and can be represented by a symmetric coset 
\begin{equation}
  \mbox{\AdS5} ~= ~ \frac{SO(2,4)}{SO(1,4)}\,. 
  \label{24/14}
\end{equation}
In fact, the metric can be computed by using (\ref{24}) and (\ref{13}) via a coset construction. 
Then a key ingredient is the left-invariant one-form 
\begin{eqnarray}
A = g^{-1}dg\,. 
\end{eqnarray}
Let us take a coset representative $g$ as \cite{AF,ABF}\footnote{
A different coset representative may be chosen. For example, see Appendix A. }
\begin{equation}
g=\exp\Bigl[\, \frac{i}{2}t\,\gamma_5-\psi_2\, n_{12}-i\psi_3\, n_{03}\,\Bigr]\,
\exp\Bigl[\,-i\psi_1\,n_{01}\,\Bigr]\,
\exp\Bigl[\,-\frac{i}{2}\rho\gamma_0\,\Bigr]\, . 
\label{global-AdS}
\end{equation}
Then the left-invariant one-form can be expanded as 
\begin{eqnarray}
A =\frac{1}{2}e^{M}\gamma_{M}  
+\frac{1}{2}\omega^{MN}n_{MN} \qquad (M, N=0,1,2,3,5)
\label{current-AdS5}
\end{eqnarray}
in terms of vielbeins $e^M$ and $\omega^{MN}$ given by 
\begin{align}
\label{global-vielbein}
e^0&=-i\,d\rho\,, &\omega^{01}&=-i\cosh\rho\, d\psi_1\,,\nonumber\\
e^1&=-\sinh\rho\, d\psi_1\,, &\omega^{02}&=-i\cosh\rho\sin\psi_1 d\psi_2\,, \nonumber\\
e^2&=-\sinh\rho\sin\psi_1 d\psi_2\,, &\omega^{03}&=-i\cosh\rho\cos\psi_1 d\psi_3\,, \nonumber\\
e^3&=-\sinh\rho\cos\psi_1 d\psi_3\,, &\omega^{05}&=-\sinh\rho\, dt\,, \nonumber\\
e^5&=i\cosh\rho\, dt\,, &\omega^{12}&=-\cos\psi_1 d\psi_2\,, \nonumber\\
&&\omega^{13}&=\sin\psi_1 d\psi_3\,,  
\end{align}
where the other components of $\omega^{MN}$ are zero. 
By using the vielbeins $e^M$\,, the global AdS$_5$ metric is given by (we set AdS radius equal to 1)
\begin{eqnarray}
ds^2 &=& g_{MN}dx^M dx^N = \eta_{MN}e^Me^N \nonumber \\ 
&=& -\cosh^2\rho\,dt^2+d\rho^2+\sinh^2\rho
\left(d\psi^2_1+\sin^2\psi_1\,d\psi^2_2+\cos^2\psi_1\,d\psi^2_3\right)\,. 
\label{GAdS}
\end{eqnarray}
Here $\eta_{MN}=\mbox{diag}(-1,1,1,1,1)~~(M,N=0,1,2,3,5)$\,. 

\medskip 

An alternative way is to adopt 
a coset projector ~$\overline{P} ~:~ \alg{so}(2,4) \to \alg{so}(2,4)/\alg{so}(1,4)$\,,
\begin{eqnarray}
\overline{P}(x)  &\equiv& \ga_{0}\,\frac{\Tr(\ga_{0}\,x)}{\Tr(\gamma_0^2)}
+\sum_{i=1}^3\ga_{i}\,\frac{\Tr(\ga_{i}\,x)}{\Tr(\gamma_i^2)}
+ \gamma_5\,\frac{\Tr(\gamma_5\,x)}{\Tr(\gamma_5^2)} \nonumber \\
&=& \frac{1}{4} \Bigl[-\ga_{0}\,\Tr(\ga_{0}\,x) + \sum_{i=1}^3\ga_{i}\, \Tr(\ga_{i}\,x)
+\gamma_5\,\Tr(\gamma_5\,x)
\Bigr] 
 \qquad \text{for}~~x \in \alg{so}(2,4)\,. \label{projector}
\end{eqnarray}
This projection respects the $\mathbb{Z}_2$-grading of the coset (\ref{24/14})\,. 
Then the AdS$_5$ metric (\ref{GAdS}) can be reproduced as 
\begin{equation}
ds^2 ={\rm Tr}(A\overline{P}(A))\,. 
\label{metric-formula}
\end{equation}

\medskip

Finally, let us comment on the relation between the calculation using 
the projection (\ref{projector}) and the standard coset construction. 
By using the relations, which are valid for our representation, 
\begin{eqnarray}
{\rm Tr}(\gamma_{M}\,\gamma_{N}) = 4\eta_{MN}\,, \quad 
{\rm Tr}(n_{MN}n_{PQ}) =\eta_{MQ}\eta_{NP}-\eta_{MP}\eta_{NQ}\,, 
\end{eqnarray}
the vielbeins $e^M$ are expressed as 
\begin{eqnarray}
e^M=\frac{1}{2}\Tr(\gamma^MA)\,. 
\label{viel-global}
\end{eqnarray}
Hence the expression (\ref{GAdS}) can be rewritten as 
\begin{eqnarray}
ds^2 &=& \eta_{MN}e^Me^N \nonumber \\
&=& \frac{1}{4}\eta_{MN}{\rm Tr}(\gamma^{M}A){\rm Tr}(\gamma^{N}A) 
={\rm Tr}\left[A~\frac{1}{4}\eta^{MN}\gamma_M{\Tr}(\gamma_{N}\,A)\right] \nonumber \\ 
&=&{\rm Tr}(A \overline{P}(A))\,. 
\end{eqnarray}
Thus the projection method is equivalent to the standard one. 

\medskip 

As a side note, it is apparent to see the relation 
between the general symmetric two-form \cite{NW,SYY} 
and the projection operator. This observation enables us to consider Yang-Baxter deformations 
of the Schr\"odinger and Lifshitz cosets argued in \cite{SYY}. 
We hope that we shall report these results in another place.

\subsection{Coset construction of Poincar\'e AdS$_5$}

Next, let us consider how to describe the AdS$_5$ space with the Poincar\'e coordinates 
(which will be abbreviated as the Poincar\'e AdS$_5$ hereafter). 

\medskip 

To move from the global coordinates to the Poincar\'e ones, 
we shall take another representative 
like\footnote{For the gauge transformation, see Appendix A.} 
\begin{eqnarray}
g_P&=&\exp\Bigl[ \,p_0\, x^0 + p_1\, x^1 + p_2\, x^2 + p_3\, x^3 \,\Bigr]
\exp\Bigl[\,\frac{\gamma_5}{2}\,\log z\,\Bigr] \nonumber\\
&=& \begin{pmatrix}
~\sqrt{z} & ~0~ & \frac{x^0 + x^3}{\sqrt{z}} & \frac{-x^1 + i x^2}{\sqrt{z}} \\ 
0 & \sqrt{z} & \frac{x^1 +i x^2}{\sqrt{z}} & \frac{-x^0 + x^3}{\sqrt{z}} \\ 
0 & 0 & \frac{1}{\sqrt{z}} & 0 \\ 
0 & 0 & 0 & \frac{1}{\sqrt{z}}
\end{pmatrix}
\,,
\label{para-AdS}
\end{eqnarray}
where $p_{\mu}$'s are defined as
\begin{eqnarray}
p_\mu\equiv\frac{1}{2}\gamma_\mu-n_{\mu 5}\,,
\end{eqnarray}
and represented in the following $4\times 4$ matrix notation:
\begin{eqnarray}
&& p_0 = \begin{pmatrix} 
\,0 & ~0~ & ~1~ & 0\, \\ 
0 & 0 & 0 & -1 \\ 
0 & 0 & 0 & 0 \\ 
0 & 0 & 0 & 0 
\end{pmatrix}
\,, \qquad 
p_1 = \begin{pmatrix} 
\,0 & ~0~ & ~0~ & -1 \\ 
0 & 0 & 1 & 0 \\ 
0 & 0 & 0 & 0 \\ 
0 & 0 & 0 & 0 
\end{pmatrix}
\,,  
\nonumber \\ && 
p_2 = \begin{pmatrix} 
~0 & ~0~ & ~0~ & i~\\ 
0 & 0 & i & 0 \\ 
0 & 0 & 0 & 0 \\ 
0 & 0 & 0 & 0 
\end{pmatrix}
\,, \qquad p_3 = \begin{pmatrix} 
~0 & ~0~ & ~1~ & 0~ \\ 
0 & 0 & 0 & 1 \\ 
0 & 0 & 0 & 0 \\ 
0 & 0 & 0 & 0 
\end{pmatrix}
\,.
\end{eqnarray}
Note here that $p_{\mu}$'s commute with each other. 

\medskip 

Then the left-invariant one-form $A$ can be expanded as 
\begin{eqnarray}
A \equiv g_P^{-1}dg_P 
= \begin{pmatrix}
~\frac{dz}{2z} & ~0~ & \frac{dx^0 + dx^3}{z} & \frac{-dx^1 + i dx^2}{z} \\ 
0 & \frac{dz}{2z} & \frac{dx^1 +i dx^2}{z} & \frac{-dx^0 + dx^3}{z} \\ 
0 & 0 & -\frac{dz}{2z} & 0 \\ 
0 & 0 & 0 & -\frac{dz}{2z}
\end{pmatrix}
=e^\mu p_\mu + e^5 \frac{\gamma_5}{2}\,,
\end{eqnarray}
where $\mu=0,1,2,3$ and the vielbeins $e^M~(M=0,1,2,3,5)$ are given by 
\begin{eqnarray}
\label{P-vielbein}
e^{\mu} = \frac{dx^{\mu}}{z}\,, \qquad e^5 = \frac{dz}{z}\,. 
\end{eqnarray}
Then the metric of the Poincar\'e AdS$_5$ is derived as 
\begin{eqnarray}
ds^2&=&g_{MN}dx^M dx^N = \eta_{MN}e^Me^N \nonumber\\
&=&\frac{1}{z^2}\left[-(dx^0)^2 + \sum_{i=1}^3(dx^i)^2 + dz^2\right]\,. \label{PAdS}
\end{eqnarray}
The AdS radius is set to 1, again. 

\medskip 

A subtle point in this construction is that a matrix representation has been used
in computing the vielbeins (\ref{P-vielbein})\,. More formally, 
those should have been derived without introducing a specific representation of $p_{\mu}$'s and $\gamma_5$\,. 

\medskip 

In order to apply Yang-Baxter deformations, it is necessary to use the trace operation 
in calculating the vielbeins. Because ${\rm Tr}(p_{\mu} p_{\nu})=0$\,, we should adopt, instead of $p_{\mu}$\,, 
some other elements of dual basis which are not orthogonal to $p_{\mu}$\,, 
in order to obtain the formulae (\ref{P-vielbein}) and (\ref{PAdS}) using techniques employed in Yang-Baxter sigma models. 
We checked that a possible choice is provided by matrices $\gamma_{\mu}$'s, which generate the curved translations on 
$SO(2,3)/SO(1,3)$\,.  
Then the $\mu$ component of the vielbein can be expressed 
as\footnote{Another choice of dual generators providing the formulae (\ref{P-vielbein}) and (\ref{PAdS}) 
is given by conformal generators $k_{\mu}$\,. We shall consider this possibility in Appendix B.},  
\begin{eqnarray}
e^{\mu} = \frac{1}{2}{\rm Tr}(\gamma^{\mu}A)\,. \label{proj-m}
\end{eqnarray}
By using this definition, the metric can be written as 
\begin{eqnarray}
ds^2 = \eta_{\mu\nu}e^{\mu}e^{\nu} = {\rm Tr}(A \overline{P}(A))\,,
\end{eqnarray}
where the projector $\overline{P}$ is defined in (\ref{projector})\,. 
As a result, the coset structure (\ref{24/14}) is still preserved in this construction. 
This point would play an important role in the next section.

\section{Yang-Baxter deformations of Minkowski spacetime}

We will introduce Yang-Baxter deformations of 
4D Minkowski spacetime by following \cite{YB-Min}.

\subsection{Coset construction of 4D Minkowski spacetime}\label{sec:conf-emb}

Let us introduce a coset construction of 4D Minkowski spacetime 
in order to perform its Yang-Baxter deformations in the next subsection.  

\medskip

To illustrate a key point, it is worth to revisit the standard coset construction 
based on the following coset: 
\begin{eqnarray}
\mbox{4D Minkowski} ~=~\frac{ISO(1,3)}{SO(1,3)}\,. 
\end{eqnarray} 
The 4D Poincar\'e algebra $\alg{iso}(1,3)$ and the 4D Lorentz algebra $\alg{so}(1,3)$ 
are spanned like 
\begin{equation}
\begin{aligned}
  \alg{iso}(1,3)&= {\rm span}_{\mathbb{R}}\{~n_{\mu\nu}\,,p_\mu~|~\mu,\nu=0,1,2,3~\}\,, \\
  \alg{so}(1,3)&= {\rm span}_{\mathbb{R}}\{~n_{\mu\nu}~|~\mu,\nu=0,1,2,3~\}\,. 
\end{aligned}
\end{equation}
A representative element $g_m$ is represented by 
\begin{equation}
g_m=\exp\Bigl[ \,p_0\, x^0 + p_1\, x^1 + p_2\, x^2 + p_3\, x^3 \,\Bigr] 
= \begin{pmatrix}
~1 & ~0~ & x^0 + x^3 & -x^1 + i x^2 \\ 
0 & 1 & x^1 +i x^2 & -x^0 + x^3 \\ 
0 & 0 & 1 & 0 \\ 
0 & 0 & 0 & 1
\end{pmatrix}
\,.
\label{para}
\end{equation}

\medskip 

Then the left-invariant one-form $A = g_m^{-1} dg_m$ is expanded as 
\begin{eqnarray}
A = 
\begin{pmatrix}
~0 & ~0~ & dx^0 + dx^3 & -dx^1 + i dx^2 \\ 
0 & 0 & dx^1 +i dx^2 & -dx^0 + dx^3 \\ 
0 & 0 & 0 & 0 \\ 
0 & 0 & 0 & 0
\end{pmatrix}
\equiv e^{\mu} p_{\mu}\,, 
\end{eqnarray}
and the vielbeins are obtained as 
\begin{eqnarray}
e^{\mu} = dx^{\mu}\,. \label{viel-P}
\end{eqnarray}
Finally, by adopting $\eta_{\mu\nu}$ to contract the vielbeins (\ref{viel-P}), 
the metric is computed as 
\begin{eqnarray}
ds^2 = \eta_{\mu\nu}e^{\mu}e^{\nu} = \eta_{\mu\nu}dx^{\mu}dx^{\nu}\,. \label{metric-P}
\end{eqnarray}
This is the standard derivation of the Minkowski metric with a coset construction. 

\medskip 

Note that the above derivation is well adjusted to our further aim of calculating Yang-Baxter deformations of 
Minkowski space. It appears that we can get also the formula (\ref{viel-P}) using the trace operation.
\begin{eqnarray}
e^{\mu} = \frac{1}{2}{\rm Tr}(\gamma^{\mu}A)\,, \label{trace-M}
\end{eqnarray}

\medskip 

We should observe that $\gamma^{\mu}$ is not contained as a generator in $\mathfrak{iso}(1,3)$ and hence performing the trace operation (\ref{trace-M}) 
is linked with the embedding 4D Minkowski space into the conformal group SO(2,4), or more explicitly, with the embedding of 4D Minkowski space 
into Poincar\'e AdS$_5$ manifold.
In particular, the coset element (\ref{para}) is obtained  
by dividing the one in (\ref{para-AdS}) 
by the dilatation (generated by $\gamma_5$)\,, 
what means that the radial coordinate $z$ has been removed. 

\medskip 

Next, by using the trace operation (\ref{trace-M}), the metric can be rewritten as 
\begin{eqnarray}
ds^2 = \eta_{\mu\nu}e^{\mu}e^{\nu} = \Tr(AP(A))\,, \label{metric-new}
\end{eqnarray}
with a new projection operator 
\begin{align}
P(x)= \frac{1}{4} \Bigl[-\ga_{0}\,\Tr(\ga_{0}\,x) + \sum_{i=1}^3\ga_{i}\, \Tr(\ga_{i}\,x) \Bigr]\,.  
\label{Proj}
\end{align}
Thus the conformal embedding enables us to compute the Minkowski metric 
by using the trace operation and to perform Yang-Baxter deformations.

\subsection{Yang-Baxter sigma model for 4D Minkowski spacetime}\label{sec:YB-4D-Min}

Yang-Baxter deformations have only been discussed for curved backgrounds so far. 
However, it is possible to apply the formulation to Minkowski spacetime. 

\medskip 

The deformed action is given by\footnote{{Here the string tension $T =\frac{1}{2\pi\alpha'}$ is set to 1, 
and the conformal gauge is taken so as to drop the dilaton coupling to 
the world-sheet scalar curvature. }}
\begin{equation}
\label{action}
S=-\frac{1}{2}\int_{-\infty}^{\infty}\!\!d\tau\int_{0}^{2\pi}\!\!d\sigma\,
(\gamma^{\alpha\beta}-\epsilon^{\alpha\beta})\, 
\Tr\Biggl[A_{\alpha}P\circ\frac{1}{1-2\eta R_{g}\circ P}(A_{\beta})\Biggr]\,, 
\end{equation}
where $A_{\alpha} = g_m^{-1}\partial_{\alpha}g_m$ and $g_m$ is given in Eq.~(\ref{para})\,. 
Here $\eta$ is a constant parameter and 
the action~(\ref{action}) is reduced to the undeformed one for $\eta=0$\,. 
The model is deformed on 2D Minkowski spacetime with the metric 
$\gamma_{\alpha\beta}={\rm diag}(-1,1)$\,.  
The anti-symmetric tensor $\epsilon^{\alpha\beta}$ 
is normalized as $\epsilon^{01}=1$\,. 
The operator $R_g$ is defined as 
\begin{equation}
R_g (X) \equiv g_m^{-1}R(g_mXg_m^{-1})g_m\,,
\label{min-R}
\end{equation}
Here a linear operator $R_g : 
D
\to \alg{so}(2,4)$ 
is a solution of the equation  
\begin{equation}
\bigl[R_g(x),R_g(y)\bigr]-R_g\left([R_g(x),y]+[x,R_g(y)]\right) 
= \omega [x,y]\,,\quad x, y\in D\,,
\label{CYBE-2}
\end{equation}
where $\omega$ is a constant parameter and $D$ is a vector space spanned 
by the set $\{\gamma^0,\gamma^1,\gamma^2,\gamma^3\}$\,.
When $\omega=0$\,, this equation is called the homogeneous CYBE. 
We get the modified CYBE (mCYBE) for $\omega \neq 0$\,. 
Note here that the domain of $R_g$ is restricted to $D$ due to the existence of the projection 
operator in the action\footnote{Note that the $R$-operator has a domain 
while the classical $r$-matrix has a support in the symmetry algebra. 
In the usual Yang-Baxter deformations, the domain is identified with the symmetry algebra. 
But we will follow a slightly different way. We argue that the presence of the operator $R_g\circ P$ 
in the action (\ref{action}) entitles us to restrict the natural domain of $R_g$ 
to the subspace $D$ on which $P$ projects. 
}.

\medskip 

The $R$-operator is related to the {\it skew-symmetric} classical $r$-matrix 
in tensorial notation through the formula
\begin{equation}
R(X)=\Tr_{2}[r(1\otimes X)]=\sum_{i}(a_{i}\Tr(b_{i}X)-b_{i}\Tr(a_{i}X))\,,  \label{rel}
\end{equation}
where the classical $r$-matrix is given by 
\begin{equation}
r=\sum_{i}a_{i}\wedge b_{i}\equiv \sum_{i}(a_{i}\otimes b_{i}-b_{i}\otimes a_{i})\,.
\end{equation}
The generators $a_{i}, b_{i}$ are elements of $\alg{so}(2,4)$\,, i.e the Yang-Baxter deformations will be investigated 
within $\alg{so}(2,4)$\,. 
In particular, when $\omega=0$, the $r$-matrix satisfies the CYBE in the tensorial notation, 
\begin{equation}
\left[r_{12},r_{13}\right] +\left[r_{12},r_{23}\right]+\left[r_{13},r_{23}\right]=0\,. \label{CYBE-1}
\end{equation}

\medskip 

Note here that one cannot see the difference between the mCYBE and the CYBE from the classical action, 
up to the $\eta$-dependent normalization factor in front of the action. 
It is relevant to the expression of Lax pairs, as we will see in section 4.

\subsubsection*{Three classes of classical $r$-matrices}

One can divide classical $\alg{so}(2,4)$ $r$-matrices into following three classes: 
\begin{enumerate}
\item[(a)] \quad $r=$ Poincar\'e $\otimes$ Poincar\'e \\ 
\quad 1. ~~ abelian ~~e.g., \quad $r \sim p_{1} \wedge p_{2}$\,, \qquad 
2. ~~ non-abelian ~~ e.g., \quad $r \sim \sum_{i=1}^3 p_{i} \wedge n_{0i}$\,, 
\item[(b)] \quad $r=$ Poincar\'e $\otimes$ non-Poincar\'e \\ 
\quad 1. ~~ abelian ~~e.g., \quad $r \sim n_{12} \wedge \gamma_5$\,, \qquad 
2. ~~ non-abelian ~~e.g., \quad $r \sim p_0 \wedge \gamma_5$\,,   
\item[(c)] \quad $r=$ non-Poincar\'e $\otimes$ non-Poincar\'e \\ 
\quad 1. ~~ abelian ~~e.g., \quad $r \sim k_{1} \wedge k_{2}$\,, \qquad 
2. ~~ non-abelian ~~e.g., \quad $r \sim k_0 \wedge \gamma_5$\,.  
\end{enumerate}
Here, given a classical $r$-matrix $r = a\otimes b$\,, the word ``abelian'' 
means that $a$ and $b$ commute with each other\footnote{In the case that summations are 
included, one needs to be careful for the definition in detail. But we will not try to dwell on it here.}. 
In the previous work \cite{YB-Min}, some examples in the classes (a)-1 and (b)-2 have been studied in detail. 
The class (a)-1 corresponds to twist procedures; the classical $r$-matrices associated with 
various kinds of Melvin twists have been identified in \cite{YB-Min}. Some examples of the class (b)-2 
have been studied as well, in particular T-duals of dS$_4$ and AdS$_4$ have been realized 
with the use of $r \sim \gamma_5 \wedge p_0$ and $r \sim \gamma_5 \wedge p_1$\,, respectively. 
For the classification of class (a) of classical $r$-matrices, see \cite{Tolstoy}.

\medskip 

Our purpose in this paper is to study the class (a)-2, in particular describing so-called $\kappa$-deformations 
of the 4D Poincar\'e algebra. This special class will be considered in detail in the next section.

\section{Deformed backgrounds from $\kappa$-Poincar\'e $r$-matrices}

In this section, we will compute the deformed metric and associated $B$-field 
by following Yang-Baxter deformations for particular non-abelian classical $r$-matrices. 
The method we follow is similar to \cite{YB-Min}.    

\medskip 

The classical $r$-matrix associated 
with the general $\kappa$-deformation of the Poincar\'e algebra is given by 
\begin{eqnarray}
r=a^\mu\,n_{\mu\nu}\wedge p^\nu\,, \label{general}
\end{eqnarray}
where $a^\mu$ is a constant four vector describing the set of deformation parameters.
This $r$-matrix satisfies the following mCYBE in the tensorial notation: 
\begin{eqnarray}
[r_{12},r_{13}]+[r_{12},r_{23}]+[r_{13},r_{23}] 
= \frac{1}{2}(a_\mu a^\mu)\,p^{\rho}\wedge n_{\rho\sigma}\wedge  p^{\sigma}\,,  
\end{eqnarray}
where $a_\mu a^\mu=\eta_{\mu\nu}a^\mu a^\nu$\,.

\medskip 

In the following we shall consider in detail the following three cases: 
\begin{enumerate}
\item \quad The standard $\kappa$-deformation: 
\qquad $a^\mu=(\frac{1}{\kappa},0,0,0)$\,, 
\item \quad The tachyonic $\kappa$-deformation: 
\qquad $a^\mu=(0,\frac{1}{\kappa},0,0)$\,, 
\item \quad The light-cone $\kappa$-deformation: 
\qquad $a^\mu=(\frac{1}{\sqrt{2}\kappa},0,0,-\frac{1}{\sqrt{2}\kappa})$\,. 
\end{enumerate}
where $\kappa$ is a deformation parameter. Note that the case 3, 
in which a vector $a^\mu$ is light-cone, namely $a_\mu a^\mu=0$\,, and
the $r$-matrix describing light-cone $\kappa$-deformation is a solution of the CYBE.

\medskip 

We shall list the resulting backgrounds for the three $\kappa$-deformations below.

\subsection{The standard $\kappa$-deformation}

It is provided by the following $r$-matrix: 
\begin{eqnarray}
r=\frac{1}{\kappa}\sum_{i=1}^3n_{0i}\wedge p_i\,. \label{standard}
\end{eqnarray}
This $r$-matrix satisfies the mCYBE in the tensorial notation and gives rise to 
the standard $\kappa$-deformation of 4D Minkowski spacetime. 

\medskip 

Recall that, due to the presence of the projection (\ref{Proj}) in the classical action, 
the domain of $R_g$ is restricted to the following domain:
\begin{equation}
x,~y ~\in~ D = \mbox{span}_{\mathbb{R}}\{\gamma^0,\gamma^1,\gamma^2,\gamma^3 \} \,, 
\label{iso}
\end{equation}
on which the Killing form is non-degenerate. This can also be regarded as 
the metric induced from the parent coset $\mathfrak{so}(2,4)/\mathfrak{so}(1,4)$\,. 
Then one can show that the mCYBE (\ref{CYBE-2}) with $\omega=1/\kappa^2$ is satisfied.
This means that $R_g$ is of non-split type 
like the classical $r$-matrix of Drinfeld-Jimbo type \cite{DJ}. 
This case has been particularly popular in the study of Yang-Baxter deformations 
based on the mCYBE \cite{Klimcik,DMV,DMV2}.

\medskip 

One can easily calculate the $R_g$-operator for the $\kappa$-deformation. 
The computation of the deformed metric 
and two-form $B$-field is straightforward. The resulting background is given by 
\begin{eqnarray}
ds^2&=& \frac{-(dx^0)^2 + dr^2}{1-\hat{\eta}^2 r^2} + r^2 (d\theta^2 + \sin^2\theta\,d\phi^2)
\,,\nonumber \\
B&=& \frac{-\hat{\eta} r}{1-\hat{\eta}^2 r^2}\,dx^0\wedge dr\,.
\end{eqnarray}
Here $\hat{\eta}=2\,\eta/\kappa$ and we have changed the coordinates as follows: 
\begin{eqnarray}
x^1 = r\cos\phi\,\sin\theta\,, \quad 
x^2 = r\sin\phi\,\sin\theta\,, \quad x^3 = r\cos\theta\,.
\end{eqnarray}
After performing a T-duality along the $x^0$-direction \cite{Hull}, 
one can obtain the background
\begin{eqnarray}
ds^2 = (dr- \hat{\eta} r\,dx^0)^2 -(dx^0)^2 + r^2 (d\theta^2 + \sin^2\theta\,d\phi^2)\,.
\end{eqnarray}
Note that the $B$-field has disappeared now. Then, by performing a coordinate transformation 
from $x^0$ to $t$\,, 
\begin{eqnarray}
x^0 = t + \frac{1}{2\hat{\eta}}\log(\hat{\eta}^2 r^2-1)\,, 
\end{eqnarray}
one can reproduce the standard metric of dS$_4$ in static coordinates, 
\begin{eqnarray}
ds^2 = -(1-\hat{\eta}^2 r^2)dt^2 + \frac{dr^2}{1-\hat{\eta}^2 r^2} 
+ r^2 (d\theta^2 + \sin^2\theta\,d\phi^2)\,.
\end{eqnarray}
Here a cosmological horizon is located at $r=1/\hat{\eta}$\,. The cosmological constant 
has been induced from the deformation parameter $\hat{\eta}$\,.

\subsection{The tachyonic $\kappa$-deformation}

Let us next consider the following $r$-matrix: 
\begin{eqnarray}
r=\frac{1}{\kappa}(-n_{10}\wedge p_0+n_{12}\wedge p_2+n_{13}\wedge p_3) \label{tachyonic}\,.
\end{eqnarray}
This $r$-matrix also satisfies the mCYBE as well as the one given by (\ref{standard})\,. 
When we change the tensorial notation to the linear $R_g$-operator description, 
there is again a subtlety related with the degenerate inner product. 
However, for the coset elements given in (\ref{iso})\,, 
the mCYBE (\ref{CYBE-2}) with $\omega=-1/\kappa^2$ is satisfied.
Again, $x$ and $y$ are the elements of $D$\,. 
This indicates that the $R_g$-operator is of split-type, which has received comparatively 
little attention. It will be relevant in constructing the associated Lax pair in the next section.

\medskip 

The associated metric and $B$-field are given by
\begin{equation}
  \begin{aligned}
    ds^2
    ={}&\frac{dt^2+(dx^1)^2}{1+\hat{\eta}^2t^2}+t^2 \cosh^2\phi d\theta^2-t^2d\phi^2\,, \\
    B 
    ={}&\frac{-\hat{\eta} t}{1+\hat{\eta}^2t^2}dt\wedge dx^1\,,
  \end{aligned}
\end{equation}
where we have introduced new coordinates $t$\,, $\theta$ and $\phi$ through 
\begin{equation}
  x^0 = t\, \sinh\phi\,,  \quad 
  x^2 = t\, \cos\theta\, \cosh\phi\,, \quad 
  x^3 = t\, \sin\theta\, \cosh\phi \,.
\end{equation}
Note here that the $B$-field can be rewritten as a total derivative. 

\medskip

As in the previous case, it is useful to perform a T-duality along the $x^1$-direction. 
Then the resulting background is given by\footnote{{
At this stage, one can see that this metric describes AdS$_4$ 
by explicitly computing the scalar curvature and the Ricci tensor.}} 
\begin{equation}
ds^2=(dt-\hat{\eta} t\,dx^1)^2+(dx^1)^2+t^2(-d\phi^2 + \cosh^2\phi\, d\theta^2)\,.
\end{equation} 
Now the $B$-field has disappeared. Let us perform a coordinate transformation,  
\begin{equation}
x^1 = y + \frac{1}{2\hat{\eta}}\log(\hat{\eta}^2t^2+1)\,.
\end{equation}
Then the resulting metric is given by 
\begin{equation}
ds^2 = (1+\hat{\eta}^2 t^2) dy^2 + \frac{dt^2}{1+\hat{\eta}^2 t^2} 
+ t^2 (-d\phi^2+\cosh^2\phi\,d\theta^2)\,.
\end{equation}
By replacing the coordinates (with a double Wick rotation) by 
\begin{equation}
y \to it\,, \quad t \to r\,, \quad \phi \to i\theta\,, \quad 
\theta \to \phi\,,
\end{equation}
one can obtain the standard metric of \AdS4 with the global coordinates
\begin{equation}
ds^2 = -(1+\hat{\eta}^2r^2)dt^2 + \frac{dr^2}{1+\hat{\eta}^2r^2} 
+ r^2 (d\theta^2 + \cos^2\theta\,d\phi^2)\,.
\end{equation}
Here $\hat{\eta}^2$ is related to the non-vanishing curvature of AdS$_4$\,.

\subsection{The light-cone $\kappa$-deformation}

The third example of classical $r$-matrix is given by  
\begin{eqnarray}
r=\frac{1}{\sqrt{2}\kappa}\Bigl[(n_{01}-n_{31})\wedge 
p_1+(n_{02}-n_{32})\wedge p_2+n_{03}\wedge(-p_0+p_3)\Bigr]\,.
\end{eqnarray}
This corresponds to the light-cone $\kappa$-deformation of Minkowski spacetime. 
This $r$-matrix, in contrast to the previous two cases, satisfies the CYBE. 
This is also the case for the associated $R_g$-operator. 

\medskip 

Then the deformed background is given by 
\begin{eqnarray}
ds^2&=&\frac{-2 dx^+ dx^-+2 \hat{\eta} ^2 x^+ r\, 
dx^+dr-\hat{\eta}^2r^2(dx^+)^2}{1-\hat{\eta} ^2 (x^+)^2}+(dr)^2+r^2(d\theta)^2\,,\nonumber \\
B&=&\frac{\hat{\eta}}{1-\hat{\eta}^2(x^+)^2}(x^+dx^+\wedge dx^--rdx^+\wedge dr)\,,
\end{eqnarray}
where the light-cone coordinates are
\begin{eqnarray}
x^\pm\equiv\frac{x^0\pm x^3}{\sqrt{2}}. 
\end{eqnarray}
By introducing a new coordinate system
\begin{eqnarray}
x^+ = \frac{1}{\hat{\eta}}\tanh(\hat{\eta} X^+)\,, \qquad x^- 
= X^-+\frac{\hat{\eta}}{2}r^2\tanh(\hat{\eta} X^+)\,, 
\end{eqnarray}
the above metric and $B$-field can be rewritten as 
\begin{eqnarray}
ds^2&=&-2 dx^+ dx^- 
-\frac{2 \hat{\eta} ^2 r^2}{\cosh^2(\hat{\eta} x^+)} (dx^+)^2+(dr)^2 
+ r^2d\theta^2\,,\nonumber \\
B&=& \frac{1}{\hat{\eta}}d\log\bigl(\cosh(\hat{\eta} x^+)\bigr)\wedge dx^- 
-\frac{1}{2}d\tanh(\hat{\eta} x^+)\wedge dr^2\,.\label{pp-wave-Min}
\end{eqnarray}
Here we have replaced $(X^+,X^-)$ with $(x^+,x^-)$ for simplicity. 
This background is just a time-dependent pp-wave background. 
The only non-vanishing component of Ricci tensor is given by 
\begin{eqnarray}
R_{++}=\frac{4 \hat{\eta} ^2}{\cosh^2(\hat{\eta} x^+)}\,,
\end{eqnarray}
and the scalar curvature vanishes. The $(++)$-component of Einstein equations of motion 
is non-trivial and it provides a light-like matter. There is some amount of literature devoted to pp-wave 
backgrounds associated with supersymmetric Yang-Mills and D-brane theories. 

\medskip 

It would be worth mentioning about the relation between the pp-wave background (\ref{pp-wave-Min}) 
and the deformed AdS$_5$ obtained with the light-cone $\kappa$-deformation $r$-matrix \cite{Stijn2}. 
By taking a slice of the deformed AdS$_5$ with a constant $z$ and 
performing appropriate coordinate transformations, 
the same pp-wave background (\ref{pp-wave-Min}) can be reproduced. 

\medskip 

Finally, it would be worth noting that the $B$-field is a total derivative. 
Hence it is anticipated that a Ramond-Ramond flux is turned on, instead of $B$-field, 
so that the resulting background satisfies the equations of motion of type IIB supergravity. 
In order to confirm this anticipation, it would be useful 
to study a slice of the deformed AdS$_5$ \cite{Stijn2} with a supercoset construction.

\section{Lax pair for the general $\kappa$-Poincar\'e deformation}

In this section, let us consider Lax pairs associated 
with the general classical $r$-matrix (\ref{general})\,. 

\medskip 

An advantage of the Yang-Baxter sigma model approach is 
that the universal expression of Lax pair can be constructed,  
independently of concrete forms of classical $r$-matrices. 
In fact, when the geometry to be deformed is given by a group manifold itself or a symmetric coset, 
the universal Lax pair has been constructed \cite{Klimcik,DMV,DMV2,KMY-Jordanian-typeIIB,MY-YBE}. 
However, in the case of Yang-Baxter deformations of 4D Minkowski spacetime, 
it has not yet been done (see however  \cite{YB-Min}). 
Therefore, our task is to deduce the universal expression 
for $a_\mu$-dependent general $\kappa$-Poincare $r$-matrix (\ref{general}).

\medskip

Although it is not straightforward\footnote{Indeed, the first two parts of (\ref{general-Lax}) can be determined 
on the analogy of abelian twists. For the detail of the derivation, see \cite{next}.}, 
we can deduce the following expression of Lax pair for the general $r$-matrix (\ref{general}) like 
\begin{eqnarray}
\mL_\pm=P_0(J_\pm)+\la^{\pm1}\left[P(J_\pm)+P'(J_\pm)\right]-a_\mu a^\mu\,\eta^2\,\la^{\pm1}
\left[P(J_\pm)-P'(J_\pm)\right]\,, \label{general-Lax}
\end{eqnarray}
in terms of the deformed current defined as
\begin{eqnarray}
J_\pm\equiv\frac{1}{1\mp2\eta R_{g}\circ P}\,A_{\pm}\,. \label{df-cur}
\end{eqnarray}
Here the projection $P(x)$ is defined in (\ref{Proj}) and we have introduced new projection 
operators $P'(x)$ and $P_0(x)$ defined as 
\begin{eqnarray}
P'(x) \equiv \sum_{\mu=0}^3n_{\mu5}\, \frac{\Tr(n_{\mu5}x)}{\Tr(n_{\mu5}n_{\mu5})}\,, \qquad 
P_0(x) \equiv \frac{1}{2}\sum_{\mu,\nu=0}^3n_{\mu\nu}\, \frac{\Tr(n_{\mu\nu}x)}{\Tr(n_{\mu\nu}n_{\mu\nu})} \,. 
\end{eqnarray}
The last term proportional to $a_{\mu}a^{\mu}$ in (\ref{general-Lax}) is added to 
the remaining familiar form. 
Given the $r$-matrix like (\ref{general})\,, 
the deformed current is written in general as follows: 
\begin{eqnarray}
J_\pm=J_\pm^\mu\, p_\mu+J_\pm^{\mu\nu}\, n_{\mu\nu}\,.
\label{component dc}
\end{eqnarray}
The first and second terms of (\ref{component dc}) depend on $\mathfrak{iso}(1,3)/\mathfrak{so}(1,3)$ 
and $\mathfrak{so}(1,3)$ generators, respectively.
To begin with, the first term of the Lax pair (\ref{general-Lax}) is projected 
by $P_0$ to $J_\pm^{\mu\nu} n_{\mu\nu}$\,.
Then the second term is projected by $P+P'$ to $J_\pm^\mu\, p_\mu$\,, 
simply because $p_\mu=\gamma_\mu/2-n_{\mu5}$\,. Finally, 
the third term is projected by $P-P'$ to $J_\pm^\mu\, k_\mu$\,, 
where $k_\mu = \gamma_\mu/2+n_{\mu5}$ is the generator for special conformal transformation. 
Thus, the last term indicates the appearance of the special conformal generators 
in the expression of the Lax pair (\ref{general-Lax}) which are present if the mCYBE is concerned with.

\medskip 

The appearance of 
the conformal algebra may also be related to the fact that 
T-duals of 4D dS and AdS spaces can be also obtained by using other classical 
$\mathfrak{so}(2,4)$ $r$-matrices 
like $r \sim \hat{d} \wedge p_0$ and $r \sim \hat{d} \wedge p_1$\,\cite{YB-Min},
where the dilatation operator $\hat{d}\,(\equiv\gamma_5/2)$ is a non-Poincar\'e conformal generator. 

\medskip 

Our main result is that we have succeeded 
in checking that the zero curvature condition of the Lax pair (\ref{general-Lax}) 
is equivalent to the equations of motion. The detailed computation is given in this section.

\subsection{Proof for the Lax pair describing general $\kappa$-deformations}

Let us confirm that the Lax pair (\ref{general-Lax}) works well 
for the general $\kappa$-Poincar\'e $r$-matrix (\ref{general}).
It is convenient to rewrite the Lax pair (\ref{general-Lax}) as
\begin{eqnarray}
\mL_\pm&=&J^n_\pm+\lambda^{\pm1}(J^p_\pm-a^2\eta^2J^{\tilde{k}}_\pm)\,,\nonumber\\
J^p_\pm&\equiv& J^{\mu}_\pm p_\mu\,,\quad J^n_\pm 
\equiv J^{\mu\nu}_\pm n_{\mu\nu}\,,\quad J^{\tilde{k}}_\pm\equiv J^{\mu}_\pm k_{\mu}\,,\quad a^2=a_\mu a^\mu\,.
\end{eqnarray}
The zero curvature condition of the Lax pair (\ref{general-Lax}) are also rewritten as
\begin{eqnarray}
\label{lax zcc}
0&=&\partial_+\mL_--\partial_-\mL_++\left[\mL_+,\mL_-\right]\nonumber\\
&=&\lambda\left(-\partial_-J^p_++[J^p_+,J^n_-]+a^2
\eta^2\partial_-J_+^{\tilde{k}}-a^2\eta^2[J_+^{\tilde{k}},J^n_-]\right)\nonumber\\
&&+\frac{1}{\lambda}\left(\partial_+J^p_-+[J^n_+,J^p_-]-a^2
\eta^2\partial_+J_-^{\tilde{k}}-a^2\eta^2[J_+^n,J^{\tilde{k}}_-]\right)\nonumber\\
&&+\partial_+J^n_--\partial_-J^n_++[J^n_+,J^n_-]-a^2\eta^2[J^p_+,J^{\tilde{k}}_-]-a^2\eta^2[J^{\tilde{k}}_+,J^p_-]\,. 
\end{eqnarray}
The coefficients of $\lambda$, $\lambda^0$, $\lambda^{-1}$ should vanish respectively, 
the zero curvature condition is equivalent to the following three equations;
\begin{eqnarray}
\partial_+J^p_-&+&[J^n_+,J^p_-]=0\,,\nonumber\\
\partial_-J^p_+&-&[J^p_+,J^n_-]=0\,,\nonumber\\
\partial_+J^n_--\partial_-J^n_+&+&[J^n_+,J^n_-]-4a^2\eta^2J^\mu_+J^\nu_-n_{\mu\nu}=0\,.
\label{eom zcc}
\end{eqnarray}
In the derivation of the third equation in (\ref{eom zcc}) from the $\lambda^0$ term in (\ref{lax zcc})\,, 
we have used the commutation relation; $[p_\mu,k_\nu]=2n_{\mu\nu}+\eta_{\mu\nu}\gamma_5$\,.
Then the remaining task is to confirm that the three equations in (\ref{eom zcc}) 
are equivalent to the equation of motion 
of the deformed system (\ref{action}) and the zero curvature condition for $A_\mu$\,.

\subsubsection*{Equations of motion}

The equations of motion following from the deformed action (\ref{action}) 
can be written in terms of deformed current $J_\pm$ as follows:\footnote{
The derivation of (\ref{eom}) is given in Appendix A of \cite{MY-YBE}. 
The expression (\ref{eom}) always holds, 
but the classical $r$-matrix (\ref{general}) was assumed for (\ref{eom-2})\,. However, as shown in \cite{next}, 
the expression (\ref{eom-2}) is eventually valid for arbitrary classical $r$-matrices of Poicar\'e $\otimes$ Poincar\'e.} 
\begin{eqnarray}
\label{eom}
\Tr \bigl[\mathcal{E}\,p_\mu \bigr]=0\,,\quad\mathcal{E} 
\equiv \partial_+P(J_-)+\partial_-P(J_+)+[J_+,P(J_-)]+[J_-,P(J_+)]\,.
\end{eqnarray}
Then, by expanding $J_{\pm}$\,, the equations in (\ref{eom}) can be rewritten as 
\begin{eqnarray}
\Tr \biggl[\bigl(\partial_+J^\rho_-\gamma_\rho+\partial_-J^\rho_+\gamma_\rho+
\left[J^\rho_+p_\rho+J^{\rho\sigma}_+n_{\rho\sigma},J^\lambda_-\gamma_\lambda\right]
+[J^\rho_-p_\rho+J^{\rho\sigma}_-n_{\rho\sigma},J^\rho_+\gamma_\rho]\bigr)p_\mu \biggr]=0\,. 
\label{eom-2}
\end{eqnarray}
In the third and fourth terms, the commutator $[\gamma_\mu,p_\nu]~(=2n_{\mu\nu}+\eta_{\mu\nu}\gamma_5)$ 
can be dropped off because the generators given by this commutator ($n_{\mu\nu}$ and $\gamma_5$) 
vanish after taking trace with $p_\mu$\,. 
For the remaining terms, $\gamma_\mu$ can be replaced 
by $p_\mu$ because the commutation relations $[\gamma_\mu,n_{\nu\rho}]$ and $[p_\mu,n_{\nu\rho}]$ 
have the same form. 
After all, the equations of motion of the $\kappa$-deformed system (\ref{action}) 
are equivalent to the following equations: 
\begin{eqnarray}
\label{eom2}
\tilde{\mathcal{E}}\equiv\partial_+J^p_-+\partial_-J^p_++[J^n_+,J^p_-]+[J^n_-,J^p_+]=0\,.
\end{eqnarray}

\subsubsection*{Zero curvature condition}

One can rewrite the zero curvature condition for $A_\pm$ in terms of $J_\pm$ like 
\begin{eqnarray}
\label{zcc}
0&=&\mathcal{Z}=\partial_+A_--\partial_-A_++\left[A_+,A_-\right]\nonumber\\
&=&\partial_+J_--\partial_-J_++\left[J_+,J_-\right]
+2\eta R_g(\mathcal{E})+4\eta^2{\rm{YBE}}_{Rg}(P(J_+),P(J_-))\,, 
\end{eqnarray}
through the relation $A_\pm=(1\mp 2\eta R_g\circ P)J_\pm$\,. 
Now, by noting the relation 
\begin{equation}
R_g(\mathcal{E}) = R_g(\tilde{\mathcal{E}})\,, 
\end{equation}
the fourth term in the last line can be rewritten as $2\eta R_g(\tilde{\mathcal{E}})$\,.
The symbol ${\rm{YBE}}_{Rg}$ in the fifth term is defined as  
\begin{eqnarray}
{\rm{YBE}}_{Rg}(X,Y) \equiv [R_g(X),R_g(Y)]-R_g([R_g(X),Y]+[X,R_g(Y)])\,.
\end{eqnarray}
With the $r$-matrix (\ref{general})\,, the expression of ${\rm{YBE}}_{Rg}$ is explicitly evaluated as 
\begin{eqnarray}
{\rm{YBE}}_{Rg}(X,Y)&=&a^2\left(\Tr[p^\mu gXg^{-1}]\Tr[n_{\mu\nu}gYg^{-1}]g^{-1}p^\nu g\right.\nonumber\\
&&\left.+\Tr[n_{\nu\mu}gXg^{-1}]\Tr[p^\mu gYg^{-1}]g^{-1}p^\nu g\right.\nonumber\\
&&\left. +\Tr[p^\mu gXg^{-1}]\Tr[p^\nu gYg^{-1}]g^{-1}n_{\nu\mu}g\right)\,, 
\end{eqnarray}
where we have omitted the subscript $m$ of $g_m$ for simplicity.
In particular, if X and Y can be expanded by $\gamma_\mu$ only 
(that is, $X=X^\mu\gamma_\mu$ and $Y=Y^\mu\gamma_\mu$)\,, 
${\rm{YBE}}_{Rg}(X,Y)$ may take a simple form: 
\begin{eqnarray}
{\rm{YBE}}_{Rg}(X,Y)=-4a^2X^\mu Y^\nu n_{\mu\nu}\,. \label{514}
\end{eqnarray}
This equation (\ref{514}) indicates that the $r$-matrix (\ref{general}) satisfies the mCYBE 
for the elements of $\mathfrak{so}(2,3)/\mathfrak{so}(1,3)$\,. 
Thus the last line of (\ref{zcc}) is recast into 
\begin{eqnarray}
\label{zcc2}
0&=&\mathcal{Z}=\partial_+J_--\partial_-J_++[J_+,J_-] 
+ 2\eta R_g (\tilde{\mathcal{E}})-4a^2J^\mu_+ J^\nu_- n_{\mu\nu}\,.
\end{eqnarray}
Note the right-hand side of (\ref{zcc2}) has the $\mathfrak{iso}(1,3)$ components only by construction.
Then it can be decomposed into the $\mathfrak{iso}(1,3)/\mathfrak{so}(1,3)$ 
and $\mathfrak{so}(1,3)$ components as follows: 
\begin{eqnarray}
\label{zcc3}
0&=&\mathcal{Z}^p\equiv\partial_+J^p_--\partial_-J^p_+ 
+ \left[J^p_+,J^n_-\right]+\left[J^n_+,J^p_-\right]+2\eta P(R_g(\tilde{\mathcal{E}}))\,,\nonumber\\
0&=&\mathcal{Z}^n\equiv\partial_+J^n_--\partial_-J^n_+ 
+ \left[J^n_+,J^n_-\right]+2\eta P_0(R_g(\tilde{\mathcal{E}}))-4a^2J^\mu_+ J^\nu_- n_{\mu\nu}\,.
\end{eqnarray}
Note that $P_0(R_g(\tilde{\mathcal{E}}))$ and $P(R_g(\tilde{\mathcal{E}}))$ 
vanish because $\tilde{\mathcal{E}}=0$\,.

\medskip 

Now we have prepared to check the equivalence between $\tilde{\mathcal{E}}=\mathcal{Z}^p=\mathcal{Z}^n=0$ 
and the equations in (\ref{eom zcc})\,. It is manifest and the correspondence is summarized in the following:
\begin{eqnarray}
\left\{
\begin{array}{c}
~~\tilde{\mathcal{E}}+\mathcal{Z}^p = 0 \\
~~\tilde{\mathcal{E}}-\mathcal{Z}^p = 0 \\
~~\mathcal{Z}^n = 0
\end{array}
\right.
\qquad \Longleftrightarrow \qquad \mbox{three equations in}~(\ref{eom zcc})\,.
\nonumber
\end{eqnarray}
Thus we have shown that the zero curvature condition 
of the Lax pair (\ref{general-Lax}) is equivalent to 
the equations of motion of the deformed system.

\subsection{Three types of $\kappa$-deformations }

In the following, we shall show the explicit forms of Lax pair for the three types of $\kappa$-deformations, 
1) the standard $\kappa$-deformation, 2) the tachyonic $\kappa$-deformation 
and 3) the light-cone $\kappa$-deformation. 

\subsubsection*{1) Lax pair for the standard $\kappa$-deformation}

First of all, for the standard $\kappa$-deformation, the deformed current is given by
\begin{eqnarray}
J_\pm&=&\biggl(r\cos\theta\cos\phi\partial_\pm\theta-r\sin\theta\sin\phi\partial_\pm\phi
+\sin\theta\cos\phi\frac{\partial_\pm r\pm\hat{\eta} r\partial_\pm x^0}{1-\hat{\eta}^2r^2}\biggr)
\left[p_1\pm \hat{\eta}\,n_{01}\right] \nonumber \\
&&+\biggl(r\cos\theta\sin\phi\partial_\pm\theta+r\sin\theta\cos\phi\partial_\pm\phi 
+ \sin\theta\sin\phi\frac{\partial_\pm r\pm\hat{\eta} r\partial_\pm x^0}{1-\hat{\eta}^2r^2}\biggr)
\left[p_2\pm \hat{\eta}\,n_{02}\right]\nonumber \\
&&+\biggl(-r\sin\theta\partial_\pm\theta
+\cos\theta\frac{\partial_\pm r\pm\hat{\eta} r\partial_\pm x^0}{1-\hat{\eta}^2r^2}\biggr)
\left[p_3\pm\hat{\eta}\, n_{03}\right] 
+\frac{\partial_\pm x^0\pm\hat{\eta} r\partial_\pm r}{1-\hat{\eta}^2r^2}p_0\,.
\end{eqnarray}
Then, by using this $J_{\pm}$\,, the explicit form of Lax pair is given by
\begin{eqnarray}
\mL_\pm&=&\biggl(r\cos\theta\cos\phi\partial_\pm\theta-r\sin\theta\sin\phi\partial_\pm\phi \nonumber \\
&&\qquad\quad\qquad+\sin\theta\cos\phi\frac{\partial_\pm r\pm\hat{\eta} r\partial_\pm x^0}{1-\hat{\eta}^2r^2}\biggr)
\left[\la^{\pm1}\left(p_1+\frac{\hat{\eta}^2}{4}k_1\right)\pm \hat{\eta}\,n_{01}\right] \nonumber \\
&&+\biggl(r\cos\theta\sin\phi\partial_\pm\theta+r\sin\theta\cos\phi\partial_\pm\phi \nonumber \\
&&\qquad\quad\qquad+\sin\theta\sin\phi\frac{\partial_\pm r\pm\hat{\eta} r\partial_\pm x^0}{1-\hat{\eta}^2r^2}\biggr)
\left[\la^{\pm1}\left(p_2+\frac{\hat{\eta}^2}{4}k_2\right)\pm \hat{\eta}\,n_{02}\right] \nonumber \\
&&+\biggl(-r\sin\theta\partial_\pm\theta+\cos\theta\frac{\partial_\pm r\pm\hat{\eta} r\partial_\pm x^0}{1-\hat{\eta}^2r^2}\biggr)
\left[\la^{\pm1}\left(p_3+\frac{\hat{\eta}^2}{4}k_3\right)\pm\hat{\eta}\, n_{03}\right] \nonumber \\
&&+\frac{\partial_\pm x^0\pm\hat{\eta} r\partial_\pm r}{1-\hat{\eta}^2r^2}\la^{\pm1}\left[p_0+\frac{\hat{\eta}^2}{4}k_0\right]\,.
\end{eqnarray}

\subsubsection*{2) Lax pair for the tachyonic $\kappa$-deformation}

In the case of the tachynic $\kappa$-deformation, the deformed current is given by 
\begin{eqnarray}
J_\pm&=&\left(t\cosh\phi\partial_\pm\phi+\frac{1}{1+\hat{\eta}^2 t^2}
\sinh\phi(\partial_\pm t\mp\hat{\eta} t\partial_\pm x^1)\right)\left[p_0\pm\hat{\eta} n_{10}\right]\nonumber\\
&&+\biggl(-t\sin\theta\cosh\phi\partial_\pm\theta+t\cos\theta\sinh\phi\partial_\pm\phi \nonumber \\ 
&& \qquad +\frac{1}{1+\hat{\eta}^2 t^2}\cos\theta\cosh\phi(\partial_\pm t\mp\hat{\eta} t\partial_\pm x^1)\biggr)
\left[p_2\pm\hat{\eta} n_{12}\right]\nonumber\\
&&+\biggl(t\cos\theta\cosh\phi\partial_\pm\theta+t\sin\theta\sinh\phi\partial_\pm\phi \nonumber \\ 
&& \qquad +\frac{1}{1+\hat{\eta}^2 t^2}\sin\theta\cosh\phi(\partial_\pm t\mp\hat{\eta} t\partial_\pm x^1)\biggr)
\left[p_3\pm\hat{\eta} n_{13}\right]\nonumber\\
&&+\frac{1}{1+\hat{\eta}^2 t^2}\left(\partial_\pm x^1\pm\hat{\eta} t\partial_\pm t\right)p_1\,.
\end{eqnarray}
Then it is easy to compute the explicit form of the Lax pair, 
\begin{eqnarray}
\mL_\pm&=&\left(t\cosh\phi\partial_\pm\phi+\frac{1}{1+\hat{\eta}^2 t^2}
\sinh\phi(\partial_\pm t\mp\hat{\eta} t\partial_\pm x^1)\right)\left[\lambda^{\pm1}
\left(p_0-\frac{\hat{\eta}^2}{4}k_0\right)\pm\hat{\eta}\,n_{10}\right]\nonumber\\
&+&\frac{1}{1+\hat{\eta}^2 t^2}\left(\partial_\pm x^1\pm\hat{\eta} t\partial_\pm t\right)\,
\lambda^{\pm1}\left[p_1+\frac{\hat{\eta}^2}{4}k_1\right]\nonumber\\
&+&\biggl(-t\sin\theta\cosh\phi\partial_\pm\theta+t\cos\theta\sinh\phi\partial_\pm\phi\nonumber\\
&&\quad+\frac{1}{1+\hat{\eta}^2 t^2}\cos\theta\cosh\phi(\partial_\pm t\mp\hat{\eta} t\partial_\pm x^1)\biggr)
\left[\lambda^{\pm1}\left(p_2-\frac{\hat{\eta}^2}{4}k_2\right)\pm\hat{\eta}\,n_{12}\right]\nonumber\\
&+&\biggl(t\cos\theta\cosh\phi\partial_\pm\theta+t\sin\theta\sinh\phi\partial_\pm\phi\nonumber\\
&&\quad+\frac{1}{1+\hat{\eta}^2 t^2}\sin\theta\cosh\phi(\partial_\pm t\mp\hat{\eta} t\partial_\pm x^1)\biggr)
\left[\lambda^{\pm1}\left(p_3-\frac{\hat{\eta}^2}{4}k_3\right)\pm\hat{\eta}\,n_{13}\right]\,.
\end{eqnarray}

\subsubsection*{3) Lax pair for the light-cone $\kappa$-deformation}

Finally, for the light-cone $\kappa$-deformation, 
the deformed current is given by 
\begin{eqnarray}
J_\pm&=&\left(\cos\theta\partial_\pm r-r\sin\theta\partial_\pm\theta
\pm\frac{\hat{\eta}\,r\cos\theta\partial_\pm x^+}{1\mp\hat{\eta}\,x^+}\right)
\left[p_1\pm\hat{\eta}\frac{n_{01}-n_{31}}{\sqrt{2}}\right]\nonumber \\
&&+\left(\sin\theta\partial_\pm r+r\cos\theta\partial_\pm\theta\pm
\frac{\hat{\eta}\,r\sin\theta\partial_\pm x^+}{1\mp\hat{\eta}\,x^+}\right)
\left[p_2\pm\hat{\eta}\frac{n_{02}-n_{32}}{\sqrt{2}}\right]\nonumber \\
&&+\frac{(1\mp\hat{\eta}\,x^+)(\partial_\pm x^-\pm\hat{\eta} r \partial_\pm r)
+\hat{\eta}^2 r^2\partial_\pm x^+}{1-\hat{\eta}^2 (x^+)^2}\frac{p_0-p_3}{\sqrt{2}} 
\nonumber \\ &&
+\frac{\partial_\pm x^+}{1\mp\hat{\eta} x^+}\left[\frac{p_0+p_3}{\sqrt{2}}\pm\hat{\eta}\,n_{03}\right]\,. 
\end{eqnarray}
Then, by using the above current $J_{\pm}$, the explicit form of Lax pair is obtained as  
\begin{eqnarray}
\mL_\pm&=&\left(\cos\theta\partial_\pm r-r\sin\theta\partial_\pm\theta\pm
\frac{\hat{\eta}\,r\cos\theta\partial_\pm x^+}{1\mp\hat{\eta}\,x^+}\right)
\left[\la^{\pm1} p_1\pm\hat{\eta}\frac{n_{01}-n_{31}}{\sqrt{2}}\right]\nonumber \\
&&+\left(\sin\theta\partial_\pm r+r\cos\theta\partial_\pm\theta
\pm\frac{\hat{\eta}\,r\sin\theta\partial_\pm x^+}{1\mp\hat{\eta}\,x^+}\right)
\left[\la^{\pm1} p_2\pm\hat{\eta}\frac{n_{02}-n_{32}}{\sqrt{2}}\right]\nonumber \\
&&+\frac{(1\mp\hat{\eta}\,x^+)(\partial_\pm x^-\pm\hat{\eta} r \partial_\pm r)
+\hat{\eta}^2 r^2\partial_\pm x^+}{1-\hat{\eta}^2 (x^+)^2}\frac{\la^{\pm1}(p_0-p_3)}{\sqrt{2}}\nonumber \\
&&+\frac{\partial_\pm x^+}{1\mp\hat{\eta} x^+}\left[\la^{\pm1}\frac{p_0+p_3}{\sqrt{2}}\pm\hat{\eta}\,n_{03}\right]\,.
\end{eqnarray}

\section{Conclusion and discussion}

In this paper, we have studied Yang-Baxter sigma models leading to deformed 4D Minkowski spacetimes 
arising from classical $r$-matrices associated with $\kappa$-deformations of the Poincar\'e algebra. 
We have considered three deformations: 
1) the standard $\kappa$-deformation, 
2) the tachyonic $\kappa$-deformation, 
and 3) the light-cone $\kappa$-deformation. 
For each of these deformations, the metric and two-form $B$-field have been computed. 
The first two deformations are related with mCYBE and lead 
to T-duals of dS$_4$ and AdS$_4$\,, respectively. 
The third deformation is linked with the CYBE and leads to a time-dependent pp-wave background. 
Finally, we have constructed a Lax pair for the generalized $\kappa$-Poincar\'e $r$-matrix. 

\medskip 

It should be mentioned that the three deformed geometries 
for the standard and tachyonic $\kappa$-deformations 
have already been obtained in \cite{YB-Min} with different kinds of classical $r$-matrices 
which are composed of $\gamma_5$ and a Poincar\'e generator. 
In addition, such a classical $r$-matrix has been found also 
for the light-cone $\kappa$-deformation and the correspondence list with the deformations of flat Minkowski space
is shown in Tab.\,\ref{list:tab}.  
It is very interesting to try to understand the fundamental mathematical structure 
behind this correspondence. A preliminary clarification is given in Appendix C. 

\begin{table}[htbp]
\vspace*{0.3cm}
\begin{center}
\begin{tabular}{lcc}
\hline
\qquad  Geometry  & The present paper & The previous work \cite{YB-Min} \\ 
 \hline \hline 
~~1) T-dual of 4D dS & the standard $\kappa$-deformation & $r\sim\gamma_5\wedge p_0$ \\ 
~~2) T-dual of 4D AdS & the tachyonic $\kappa$-deformation &  $r\sim\gamma_5\wedge p_1$  \\
~~3) time-dependent pp-wave & the light-cone $\kappa$-deformation &  $r\sim\gamma_5\wedge (p_0-p_1)$ \\ 
\hline
\end{tabular}
\caption{The correspondence list of 4D (A)dS and classical $r$-matrices.  \label{list:tab}}
\end{center}
\vspace*{-0.7cm}
\end{table}

\medskip 

In this paper, we have constructed the Lax pair for the general $a_\mu$-dependent $\kappa$-deformation. 
It should be of great significance to further extend this Lax pair for general classical $r$-matrices 
beyond the $\kappa$-deformation, especially for the ones belonging to $\alg{so}(2,4)$ but including 
non-Poincar\'e generators. 

\medskip 

There are many open problems. 
It would be interesting to study soliton solutions 
by employing the classical inverse scattering method with the Lax pair constructed here. 
There may be some potential applications of the solutions in the study of string theory, 
especially in AdS/CFT dualities. So far, we have constructed Lax pairs. 
It is also nice to pursue the complete integrability by explicitly constructing the action-angle variables.

\medskip

Then it is important to study the symmetry algebras associated with the Yang-Baxter deformations. 
It would also be interesting to consider supersymmetric extensions of our result. 
Along this line, the preceding works \cite{super,Borowiec:2015rty} would be useful. 
It is also important to present our results extended to the quantized (super)string level. 

\medskip 

We hope that our results could as well shed light on some new aspects of classical and quantum gravity.

\subsection*{Acknowledgments}

We are very grateful to Takuya Matsumoto, Anna Pachol 
and Stijn J.~van Tongeren for useful discussions. 
The work of A.B. and J.L. has been supported 
by Polish National Science Center project 2014/02/B/ST2/04043, 
2013/09/B/ST2/03435 and by COST (European Cooperation in Science and Technology) 
Action MP1405 QSPACE.  
The work of K.Y. is supported by Supporting Program for Interaction-based Initiative Team Studies 
(SPIRITS) from Kyoto University and by the JSPS Grant-in-Aid for Scientific Research (C) No.15K05051.
This work is also supported in part by the JSPS Japan-Russia Research Cooperative Program 
and the JSPS Japan-Hungary Research Cooperative Program.

\appendix 

\section*{Appendix}

\section{A connection among coset representatives of AdS$_5$}

In section 2, we have used a coset representative given in (\ref{global-AdS})\,. 
As a matter of course, one may take another one. 

\medskip 

For example, let us consider the following representative \cite{AF}~:
\begin{eqnarray}
\hat{g}=\exp\left(\frac{i}{2}t\,\gamma_5\right)\,
\frac{\mathbf{1}+\frac{1}{2}(y^i\gamma_i+i y^0 \gamma_0)}{\sqrt{1-y^2/4}} 
\qquad (y^2=y^iy^i+y^0y^0\,, ~i=1,2,3)\,. 
\label{AF-g}
\end{eqnarray}
Then the AdS$_5$ metric is expressed as 
\begin{eqnarray}
ds^2=-\left(\frac{1+y^2/4}{1-y^2/4}\right)^2dt^2
+\frac{1}{(1-y^2/4)^2}\left(dy^2+y^2d\Omega^2_3\right)\,.
\end{eqnarray}

In general, different choices of coset representatives are related each other 
through coordinate transformations.
In fact, the two representatives $g$ in (\ref{global-AdS}) and $\hat{g}$ in (\ref{AF-g}) 
are related by the following gauge transformation
\begin{eqnarray}
\hat{g}=g\,h^{-1}\,,\quad h=\exp\left(-\psi_2 n_{12}\right)
\exp\left(-i\psi_3 n_{03}\right)\exp\left(-i\psi_1 n_{01}\right)
\end{eqnarray}
and the following coordinate transformation
\begin{eqnarray}
y^0&=&-2\tanh\frac{1}{2}\rho\,\cos\psi_1\,\cos\psi_3\,, \qquad 
y^1 = -2\tanh\frac{1}{2}\rho\,\sin\psi_1\,\cos\psi_2 \,,\nonumber \\
~y^2&=&-2\tanh\frac{1}{2}\rho\,\sin\psi_1\,\sin\psi_2\,,\qquad
y^3=-2\tanh\frac{1}{2}\rho\,\cos\psi_1\,\sin\psi_3\,.
\nonumber
\end{eqnarray}

Furthermore, the representatives $g_P$ in (\ref{PAdS}) and $\hat{g}$ in (\ref{AF-g})
are related by the following gauge transformation
\begin{eqnarray}
g_P=\hat{g}\,{h'}^{-1}\,,\quad h'=\frac{1}{\sqrt{1+y^2/4}}\left(\mathbf{1}+y^in_{i5}+i y^0 n_{05}\right)\,,\quad (i=1,2,3)
\end{eqnarray}
and coordinate transformation
\begin{eqnarray}
x^0&=&\frac{i}{1+y^2/4}\,e^{it}\,y^0\,,
\quad x^1=\frac{1}{1+y^2/4}\,e^{it}\,y^1 \,,
\quad x^2=\frac{1}{1+y^2/4}\,e^{it}\,y^2\,,\nonumber\\
\quad x^3&=&\frac{1}{1+y^2/4}\,e^{it}\,y^3\,,
\quad z=\frac{1-y^2/4}{1+y^2/4}\,e^{it}\,.
\end{eqnarray}
Thus $g$, $g_P$ and $\hat{g}$ are related by a chain of gauge transformations like
\begin{eqnarray}
g = \hat{g} h = g_{P}h' h\,, \qquad h\,,h'\in SO(1,4)\,.
\end{eqnarray}
That is, these three group elements can be used as the locally equivalent 
representatives of the coset (\ref{24/14})\,.

\section{Another coset for the Poincar\'e AdS$_5$}

As for the Poincar\'e AdS$_5$\,, one may take another choice of the projector 
\begin{eqnarray}
\widetilde{P}(x)  &\equiv& k_{0}\,\frac{\Tr(k_{0}\,x)}{\Tr(k_0p_0)}
+\sum_{i=1}^3k_{i}\,\frac{\Tr(k_{i}\,x)}{\Tr(k_ip_i)}
+ \gamma_5\,\frac{\Tr(\gamma_5\,x)}{\Tr(\gamma_5^2)} \nonumber \\
&=& \frac{1}{2} \Bigl[-k_{0}\,\Tr(k_{0}\,x) + \sum_{i=1}^3k_{i}\, \Tr(k_{i}\,x)
+\frac{1}{2}\gamma_5\,\Tr(\gamma_5\,x)
\Bigr] 
 \qquad \text{for}~~x \in \alg{so}(2,4)\,. \label{projector-P}
\end{eqnarray}
Here $k_{\mu}$ satisfying the relation $\Tr(k_\mu p_\nu)=2\eta_{\mu\nu}$ is employed instead of $\gamma_{\mu}$\,. 
This replacement modifies the definition of vielbeins like  
\begin{eqnarray}
e^{\mu} = {\rm Tr} \left(k^{\mu}A\right)\,, \quad e^5 = \frac{1}{2} {\rm Tr}(\gamma_5 A)\,,
\end{eqnarray}
but the resulting vielbeins are the same as the ones in (\ref{P-vielbein})\,.  
Thus, by taking the same representative $g$ as (\ref{para-AdS}), 
the Poincar\'e AdS$_5$ metric can be obtained as 
\begin{eqnarray}
ds^2= \eta_{MN}e^M e^N={\rm Tr}(A\widetilde{P}(A))\,. 
\end{eqnarray}
Thus the same result has been reproduced even by using this different projection (\ref{projector-P})\,. 

\medskip 

It should be remarked that the projection (\ref{projector-P}) does not preserve 
the standard coset structure (\ref{24/14}) any more. 
Instead, it respects another coset structure,   
\begin{eqnarray}
\frac{SO(2,4)}{\widetilde{ISO}(1,3)}\,. 
\label{24/i13}
\end{eqnarray}
Here $\widetilde{ISO}(1,3)$ describes a nonstandard Poincar\'e group 
with the special conformal generators $k_{\mu}$ and the following Lie algebra
\begin{eqnarray}
\widetilde{\mathfrak{iso}}(1,3) = {\rm span}_{\mathbb{R}}\{~
k_{\mu}\,, n_{\mu\nu}~|~~\mu,\nu=0,1,2,3~\}\,. \label{13-tilde}
\end{eqnarray}
Due to the definition of the new projector (\ref{projector-P})\,, this coset (\ref{24/i13}) enjoys 
the invariance under the following gauge transformations
\begin{eqnarray}
g\rightarrow gh\,, \qquad h\in \widetilde{ISO}(1,3)\,.
\label{gauge-trf}
\end{eqnarray}

The coset (\ref{24/i13}) is not globally isomorphic via the gauge transformation (\ref{gauge-trf}) with the coset (\ref{24/14}), 
but these two cosets can be shown to be locally equivalent. 
Further, the coset (\ref{24/i13}), similarly like the one parametrized by (\ref{para-AdS}), is not symmetric.
It is interesting to consider the YB deformation corresponding to 
the coset (\ref{24/i13}) and the existence of Lax pair, what we plan to follow in our future work.

\section{The coincidence of target geometries obtained from different classical $r$-matrices}

It is significant to clarify the origin of the coincidence of geometries obtained from different classical $r$-matrices. 
For this purpose, we consider a particular pair of Yang-Baxter deformations.  
It will be helpful to see the explicit expressions of deformed current $J_{\pm}$\,. 

\medskip 

For simplicity, we will focus upon the standard $\kappa$-deformation 
\begin{eqnarray}
r^{(s)}=\frac{1}{\kappa}\sum_{i=1}^3n_{0i}\wedge p_i \label{app1}
\end{eqnarray}
and the classical $r$-matrix including the dilatation 
\begin{equation}
r^{(d)}= p_0\wedge \frac{1}{2} \hat{d}\,. \label{app2}
\end{equation}
Then the deformed currents following from the formula (\ref{df-cur}) are given by, respectively, 
\begin{eqnarray}
J^{(s)}_\pm&=&\biggl(r\cos\theta\cos\phi\partial_\pm\theta-r\sin\theta\sin\phi\partial_\pm\phi
+\sin\theta\cos\phi\frac{\partial_\pm r\pm\hat{\eta} r\partial_\pm x^0}{1-\hat{\eta}^2r^2}\biggr)
\left[p_1\pm \hat{\eta}\,n_{01}\right] \nonumber \\
&&+\biggl(r\cos\theta\sin\phi\partial_\pm\theta+r\sin\theta\cos\phi\partial_\pm\phi 
+ \sin\theta\sin\phi\frac{\partial_\pm r\pm\hat{\eta} r\partial_\pm x^0}{1-\hat{\eta}^2r^2}\biggr)
\left[p_2\pm \hat{\eta}\,n_{02}\right]\nonumber \\
&&+\biggl(-r\sin\theta\partial_\pm\theta
+\cos\theta\frac{\partial_\pm r\pm\hat{\eta} r\partial_\pm x^0}{1-\hat{\eta}^2r^2}\biggr)
\left[p_3\pm\hat{\eta}\, n_{03}\right] 
+\frac{\partial_\pm x^0\pm\hat{\eta} r\partial_\pm r}{1-\hat{\eta}^2r^2}p_0 
\end{eqnarray}
and 
\begin{eqnarray}
J^{(d)}_\pm&=&\biggl(r\cos\theta\cos\phi\partial_\pm\theta-r\sin\theta\sin\phi\partial_\pm\phi
+\sin\theta\cos\phi\frac{\partial_\pm r\pm\eta r\partial_\pm x^0}{1-\eta^2r^2}\biggr)
p_1\nonumber \\
&&+\biggl(r\cos\theta\sin\phi\partial_\pm\theta+r\sin\theta\cos\phi\partial_\pm\phi 
+ \sin\theta\sin\phi\frac{\partial_\pm r\pm\eta r\partial_\pm x^0}{1-\eta^2r^2}\biggr)
p_2\nonumber \\
&&+\biggl(-r\sin\theta\partial_\pm\theta
+\cos\theta\frac{\partial_\pm r\pm\eta r\partial_\pm x^0}{1-\eta^2r^2}\biggr)
p_3
+\frac{\partial_\pm x^0\pm\eta r\partial_\pm r}{1-\eta^2r^2}
\left[p_0\pm\eta\hat{d} \right]\,.
\end{eqnarray}

\medskip 

Now it is easy to check that the difference between the two currents $J^{(s)}$ and $J^{(d)}$ 
vanishes under the projection (\ref{Proj})\,, i.e.,  $J^{(s)}$ and $J^{(d)}$ are equivalent under the projection. 
Thus the resulting geometry and symmetry algebra become identical. 

\medskip 

It is still interesting to consider the criteria selecting currents which are equivalent under the projection. 
It should be stressed here the basic role of the projection operator, i.e., the coset choice. 
These are future problems for further studies.

\end{document}